# Long range phase coherencein double barrier magnetic tunnel junctions with largethick metallic quantum well

B. S. Tao[1,2], H. X. Yang[3],Y. L. Zuo[2], X. Devaux[2], G. Lengaigne[2], M. Hehn[2], D. Lacour[2], S. Andrieu[2], M. Chshiev[3], T. Hauet[2], F. Montaigne[2], S. Mangin[2], X. F. Han[1b*], Y. Lu[2a*]

[1]Beijing National Laboratory of Condensed Matter Physics, Institute of Physics, Chinese Academy of Sciences, Beijing 100190, China

[2]Institut Jean Lamour, UMR 7198, CNRS-Nancy Université, BP 239, 54506 Vandœuvre, France

[3]Univ. Grenoble Alpes, INAC-SPINTEC, F-38000 Grenoble, France; CEA, INAC-SPINTEC, F-38000 Grenoble, France and CNRS, SPINTEC, F-38000 Grenoble, France

*Corresponding authors\*: a) yuan.lu@univ-lorraine.fr, b) xfhan@iphy.ac.cn*

Double barrier heterostructures are model systems for the study of electron tunneling and discrete energy levels in a quantum well (QW). Until now resonant tunneling phenomena in metallicQW have been observed for limited thicknesses (1-2 nm) under which electron phase coherence is conserved. In the present study we show evidence of QW resonance states in Fe QW up to12 nmthick and at room temperature in fully epitaxial doubleMgAlO$_x$barrier magnetic tunnel junctions. The electron phase coherence displayed in this QWis of unprecedented quality because ofa homogenous interface phase shift due to the small lattice mismatch at the Fe/MgAlO$_x$ interface. The physical understanding of the critical role of interface strain on QW phase coherence will greatly promote the development of the spin-dependent quantum resonant tunneling applications.

Resonant tunneling in double barrier quantum well (QW) structures has been extensively studied because of its importance in the field of nanoelectronic science and technology[1]. The double barrier structure behaves as an optical interferometer. Thus, in order to observe electron resonant tunneling, the electron phase should be kept coherent when reflecting between the two potential barriers. Due to the loss of electron phase coherence, the resonant tunneling cannot survive if the two barriers are too far from each other. The decoherence process can be introduced by interface roughness[2], or inelastic scattering in the QW and at the interface[3], which can absorb and re-emit the electrons who lose their phase information.

Recently, the combination of spin-dependent tunneling magnetoresistance (TMR) effect[4,5] with resonant tunneling through metallic QW states[6,7] in magnetic tunnel junctions (MTJs) has triggered considerable interest in new functionalities of spintronic devices operating in the quantum tunneling regime. In these structures, the QW potential barrier can be formed either by metallic layer using the symmetry dependent band structure[8-13], or by a double oxide tunneling barriers with a much greater barrier height for better electron confinement[14-16]. As so far, to preserve a good phase coherence, the metallic QW thickness in double barrier magnetic tunnel junctions (DMTJs) has been limited to around 1-2nm. In this case, it is impossible to modulate the Fermi level energy ($E_F$) by making a direct electrical connection with the middle QW layer. To achieve good phase coherence and enhance the QW resonant tunneling effect with large well thickness, the dephasing mechanisms involved in the QW and at the metal/oxide interface must be clarified.

In this study, we provide the experimental evidence that these QW states can be greatly improved in fully epitaxial DMTJs based on spinel MgAlO$_x$ oxide barrier. This material was recently proposed as a promising barrier in MTJs[17-23] because of its non-deliquescence and small lattice mismatch with typical bcc ferromagnetic materials and Heusler alloys. Remarkably, up to 10 separated QW resonance states were observed on large size patterned MTJs (100-900μm$^2$) with 12nm thick Fe QW. Moreover, we observed that the conductance oscillation amplitude in MgAlO$_x$-based DMTJs is enhanced by almost one order of magnitude compared to MgO-based DMTJs with the same QW thicknesses. The unprecedented high quality of electron phase coherence in the QW with double MgAlO$_x$ barriers was explained in terms of the strain related interface dephasing mechanism. Such peculiar transport properties give us the possibility to directly connect with the QW layer to independently control the energy of electrons which are injected into the QW as if in a three-terminal device[24]. This added level of control could be used to generate new functionality in spin-dependent quantum resonant tunneling applications.

The epitaxial DMTJs with double MgAlO$_x$ barriers were grown by molecular beam epitaxy (MBE). Fig.1 shows the schematic sample structure and band energy profile. The QW structure is formed in the middle Fe layer sandwiched by a thin (Barrier *I*) and thick barrier (Barrier *II*). Two series of samples were fabricated on MgO (001) substrates. The structures of the first series samples are: Fe(45)/MgAlO$_x$(3ML)/Fe(*t*)/MgAlO$_x$(12ML)/Fe(10)/Co(20)/Au(15), where the thickness of the middle Fe layer were 6.3nm, 7.5nm and 12.6nm, respectively. The second series samples were designed to investigate the effect of barrier interface and the MTJ stacks are composed of: Fe(45)/Barrier *I* (3ML)/Fe(10)/Barrier *II* (12ML)/Fe(10)/Co(20)/Au(15), where both Barrier *I* and *II* can either be MgO or MgAlO$_x$. All numbers indicate thickness in nanometers and ML stands for atomic mono-layer. Please see in supplementary material for details of growth and sample structural characterization [25].

First let us evaluate the magneto-transport properties in one MgAlO$_x$ DMTJ with QW thickness *t*=7.5nm. The inset of Fig.2(a) shows the representative TMR curve measured at different temperatures from 16K to 295K. The magnetic field (*H*) was applied along the Fe[100] easy axis direction and the TMR ratio is calculated as $(R_{AP}-R_P)/R_P\times 100\%$, where $R_{AP}$ and $R_P$ are the resistance of the antiparallel (AP) and parallel (P) magnetization configurations, respectively. From the shape of TMR curves, it is possible to identify that the hard layer is the top Fe/Co layer and the free layer is composed of the middle and bottom Fe layers which are ferromagnetic coupled through the 3ML thin barrier (see more magnetic characterization in supplementary information [25]). Fig. 2(a)-(b) show the differential conductance ($dI/dV$) curves measured at different temperature in P and AP states, which are normalized with the conductance $G_P$(0V,16K). In the P state, some strong oscillations of the conductance are observed in the negative bias region where the electrons are injected from top electrodes to the QW. The separation of the local maximum peak gradually increases at higher negative bias. Although the oscillation amplitude attenuates with increasing temperature, the oscillatory feature is still observable even at RT with the periodicity remaining almost unchanged. Another interesting aspect is that the oscillatory feature is still observable in AP state although with significantly reduced amplitude. The periodicity and maximum peak positions are almost the same as those in P state, which was not observed in MgO DMTJs[14,15]. In Fig.2(c), the differential MR (DTMR) curves calculated from the differential $dI/dV$ curves in P and AP states also show clear oscillations with the same phases as those of the conductivity oscillations.

In order to confirm that the observed conductance oscillations originate from the QW states in the middle Fe layer, the bias dependent conductance was measured for two other MgAlO$_x$ DMTJs with different middle Fe thickness: 6.3nm and 12.6nm. To precisely measure the QW energy position, the secondary differential $d^2I/dV^2$ curves were deduced from the normalized $dI/dV$ curves in P state, which is plotted in Fig.2(d). The QW energy position is defined at the local minimum of the $d^2I/dV^2$ curves, as marked with the dashed lines. All samples show clear oscillation behaviors with different amplitudes and periodicity. The increase of Fe thickness results in a shorter periodicity, which proves that the observed oscillations are coming from the QW states in the middle Fe layer. The highest oscillation amplitude was observed for the Fe QW thickness $t$=7.5nm. These oscillations are surprising in both their clarity and number (up to 10) that are still observable for DMTJ with $t$=12.6nm. Since the QW state is formed in Fe majority $\Delta_1$ band[7,14], the decrease of oscillation amplitude in $t$=12.6nm sample can be understood due to the finite mean free path of majority $\Delta_1$ electron for conserving its energy, symmetry and phase information. A well-defined quantum-statistical calculation[2,32] has shown that the QW states are quenched far before the QW thickness reaches the distance of mean free path, therefore we can conclude that the mean free path as well as the phase coherence length in our Fe QW should be much longer than 12nm and the reported values[33-35]. To precisely determine this phase coherence length, MgAlO$_x$DMTJs with thicker QW are needed. This is the first time that QW resonant tunneling phenomenon is observed in metallic QWdevices of large thicknesses, at least greater than 10nm.We believe this is reasonable because that the QW states have been observedby angle-resolved photoemission measurementsin quite thick Ag layer up to 112 monolayers (24.3nm) on Fe (100) [36].As For the sample with $t$=6.3nm, the smaller amplitude could be due to the increase of middle Fe roughness after annealing due to the poor wetting property of thin metal film on oxide. (See more information for the three samples in supplementary material [25].)

The QW states obtained from experimental results can bedirectly and qualitatively compared with a simple phase accumulation model (PAM) [37]. The PAM describes thequantization condition for the existence of a QW state as:

$$2k_\perp d - \Phi_1 - \Phi_2 - \Phi_{inf} = 2\pi n \quad (1)$$

where $k_\perp = \sqrt{2m^*(E-E_L)}/\hbar$ is the crystal momentumwave vector in the film perpendicular to theinterface,$d$ is the Fe QW thickness, and $\Phi_1 = \Phi_2 = 2\sin^{-1}\sqrt{(E-E_L)/(E_U-E_L)} - \pi$ is the reflection phase shift at the two Fe/MgAlO$_x$interfaces. Furthermore, $m^*$is the effective mass of majority$\Delta_1$ electron in Fe, and $E_L$ and $E_U$ are theenergies of the lower and upper edges of the barrierband gap.Here, we set $E_L$=-1.0eVand$E_U$=3.9eV similar to Ref.[7]. An important parameter $\Phi_{inf}$is taken account for the additional phase shift at interfaces due to other effects such as interface roughness, chemical disorder, impurities and strain inhomogeneity, *etc.*.To qualitatively compare our results with such an analysis, we first set $\Phi_{inf}$=0, which will be discussed below. As displayed in Fig.2(e) in solid lines, the simulated results show fairly good agreement with the experimental results when choosing $m^*$=1. It is found that the PAM can qualitatively reproduce the QW positions for samples with 6.3 and 7.5 nm Fe. For 12.6nm Fe, larger error develops at higher bias. To further quantitatively determine the QW position with such thick Fe layer, *ab initio* calculation was performed to calculate the *s*-resolved partial DOS at the $\bar{\Gamma}$ point within the two central Fe layers in bcc Fe|[MgO]$_7$|Fe|[MgO]$_7$ structure (see supplementary material for details [25]). In the calculation, 7ML of MgO was used instead of the MgAl$_2$O$_4$ structure to simplify the calculation load and two different Fe thicknesses equal to 67ML (9.4nm) and 47ML (6.5nm) were calculated. The QW peak positions derived from the sharp majority DOS spikes of the middle Fe film are marked in Fig.2(e) with open squares. The calculated QW states prettily agree with the PAM simulations as well as experimental results, which further confirms that the observed oscillation in rather thick Fe layer is originated from the QW resonant states.

The observed long range phase coherence in such thick Fe QW in our MgAlO$_x$ DMTJs could be due to a better quality of Fe QW, Fe/MgAlO$_x$ interface or MgAlO$_x$ barrier itself. To clarify the origin, four DMTJ samples were prepared with different configurations of two types of barriers: MgO and MgAlO$_x$, as listed in Table1. Fig.3(a) and 3(b) show the normalized $d^2I/dV^2$ curves in P and AP states for two samples with thin MgO barrier (B and D), respectively. [Samples A and C are shown in supplementary material Fig. S7(a) and S7(c), respectively] A clear feature is that the samples with thick MgAlO$_x$ barriers (C and D) have almost one order stronger oscillation amplitude in the P state than those with thick MgO barriers (A and B). It seems that the bottom thin barrier has no influence on the oscillation amplitude, regardless of whether it is MgO or MgAlO$_x$. In the AP state, no oscillatory feature can be observed for the samples with thick MgO barrier (A and B). The two peaks at -0.2eV and -1.0eV are related to the interface resonant state at high quality Fe/MgO interface[38]. However, for samples with thick MgAlO$_x$ barrier (C and D), clear but attenuated oscillation is still maintained in the AP state with the same periodicity as the P state. This can be understood by the band folding effect in spinel MgAl$_2$O$_4$ MTJ, as demonstrated by the *ab-initio* calculation[23]. This band folding effect induces a coupling of the $\Delta_1$ evanescent state inside the barrier with the minority-spin state in the Fe electrode, which enhances the $\Delta_1$ conductance in AP state and results in the observed resonant tunneling oscillation. We schematically illustrate this mechanism in the insets of Fig.3(a) and (b) by taking into account of only the $\Delta_1$ conductance. When the band folding effect is present, the additional $\Delta_1$ conductance in AP state will also limit the TMR ratio[18,23]. As exactly found in Table 1, the TMR ratios in DMTJs with thick MgAlO$_x$ barriers (<200% at 16K) are lower than those with thick MgO barriers (~300% at 16K).

Since the samples have the same bottom stack layers except the thick barrier[for (A and C) or (B and D)], the quality of their Fe QW should be identical. This excludes the possibility of different QW quality as the reason for the different oscillation amplitude. In addition, the contribution of the majority $\Delta_1$ channel in total parallel conductance $(G_{\Delta 1}^{\uparrow\uparrow}/G_P)$ was examined to determine if the symmetry filtering effect of the barrier[39] could play a role. However, the slight difference of $(G_{\Delta 1}^{\uparrow\uparrow}/G_P)$ in both types of barriers estimated from their TMR ratio cannot explain the one order higher oscillation amplitude in MgAlO$_x$ DMTJs (see supplementary material for details [25]). Finally, all evidence points to the Fe/thick barrier interface. To elucidate the interface related mechanisms, we have calculated electrostatic potential profiles for two structures: Fe11|[MgO]5 and Fe11|[MgAl$_2$O$_4$], as shown in Fig.3(c). Since the wavefunction decays exponentially inside the barrier region, sufficient barrier height is important to confine the QW states. The first requirement is to check the barrier heights for both MgO and MgAl$_2$O$_4$. It is found that the bond of Fe-O is stronger at Fe/MgAl$_2$O$_4$ interface (Fe-O distance 2.03Å) than that of Fe/MgO interface (2.13Å). This induces a slightly higher potential barrier at Fe/MgAl$_2$O$_4$ interface. However, the potential barrier decreases within the insulator. Therefore the Fe/MgAl$_2$O$_4$ interface has no advantage from the comparison of the barrier height. Then the lattice constant of Fe is examined.Due to the large lattice mismatch between Fe and MgO, it is clear that the Fe is under a tensile strain at the interface with MgO since the lattice constant is 2.96Å in plane and 2.72Å out-of-plane. For the $MgAl_2O_4$ case,the strain in Fe is much smaller with a lattice constant of 2.88Å in plane and 2.78Å out-of-plane. Due to the interface strain, the period of electrostatic potential in Fe is changed. It can be seen in Fig.3(c)that in the case of theFe/MgO interface, the variation of potential valley width for the interfacial Fe layers are much stronger (from 1.10Å to 1.46Å) compared to those at Fe/$MgAl_2O_4$interface (from 1.30Å to 1.45Å). This condition is even valid farther from the interface, where the Fe potential valley width varies from 1.39Å to 1.34Å in Fe/MgOcase while only from 1.39Å to 1.38Å in Fe/$MgAl_2O_4$case.These irregular potential period changes at Fe/MgO interface will undoubtedly result in a significant interface phase shift$\Phi_{inf}$for the QW states, as illustrated in Eq. (1). However, since the collected current is from atwo-dimensional spatial integration in the whole junction area, if the electron changesits phase with the same$\Phi_{inf}$everywhere, strong QW oscillationscan still be obtained but with a shift of energy positions. Therefore, other mechanismsshould exist toinduce a large distribution of $\Phi_{inf}$causing the vanishing of QW oscillation. As schematically shown in Fig.3(d), with a large distribution of$\Phi_{inf}$, the QW energy position will also have a large distributionwithin the same QW indexat different spatial locations. As a consequence, this smears the contrast of current intensity as function ofbias and gives rise tothe decrease of QW oscillation amplitude.

The mechanism introducing a large$\Phi_{inf}$distribution can be highlighted by the creation of misfit dislocation due to the lattice mismatch induced interface strain. For the Fe/MgO case, when the thickness of MgO goes beyond 5ML [40], 1/2<011> misfit dislocations occur to relax the MgO lattice which can then propagate to the bottom interface.These are observed by noting the appearance of a "V" shapedfeature [41-43]in the RHEED pattern [Supplementary Fig.S1(b)].Thispropagation will result in a long range stress field around core dislocations [44] leading to significant interatomic distance distribution in the plane of the interface. Such a distribution generatesan important phase shift scattering by varying the potential period as shown in *ab-initio* calculation andconsequently leads to a large$\Phi_{inf}$spreading.The average distance *L* between the adjacent misfit dislocations can be estimated to be 6nm for MgO and 90nm for MgAl$_2$O$_4$from a crude static model [44]:*L*=*a*/2*f*where *a*isthe lattice constant of the oxide and *f* is the mismatch with Fe.Important to note is that at the Fe/thin oxide (3ML) interface, the lattice of MgOis unrelaxed without misfit dislocation creation as confirmed bythe RHEED pattern[Supplementary Fig.S1(a)]. No important difference in conductivity oscillation amplitude is foundwhen bottom thin barrier (3ML) is MgO or MgAlO$_x$.For the 12ML thick MgAlO$_x$ layer on Fe, the small lattice mismatch induced strain creates very few misfit dislocations. In the RHEED patterns [Supplementary Fig.S1(b)], the dislocation related "V" shape feature is not observed. Furthermore, the persistence of Kikuchi lines signifies a good crystalline coherence in the whole barrier. Finally, the elimination of misfit dislocation conserves a homogenous and small distribution of$\Phi_{inf}$, and in turn results in a significant enhancement ofQW conductivity oscillation.

In summary, the spin-dependent resonant tunneling properties in fully epitaxial MgAlO$_x$DMTJs grown by MBEhave been studied. QW statesin conductance curves as a function of the bias voltage are evidenced for up toa 12nm thick Fe QW layer. Both PAM simulations and first principle calculations agree well with experimental results. Comparingexperimental results using either MgO orMgAlO$_x$insulating barriers in these DMTJs allows us to highlight the key role of misfit dislocations in the barriers for the QW state establishment. Significant enhancement of the amplitude of the conductivity oscillation is observed up to one order in the MgAlO$_x$ DMTJ. This illustrates that the control of interface strain is essential to the preservation of a homogenous interface phase shift in order to obtain a sizable QW resonant tunneling oscillation.

**Acknowledgments**

This work is supported by joint France-China (ANR-NSFC) SISTER (ANR-11-IS10-0001) and ENSEMBLE (ANR-14-CE26-0028-01) projectsas well as Région Lorraine. X. F. Han acknowledges Ministry of Science and Technology (MOST) projects (No. 2010CB934401 and No. 2011YQ120053) and the National Natural Science Foundation (NSFC, Grant No. 11434014).

**Figure captions:**

FIG. 1(Color on line).(a)Energy profile in DMTJ structure and QW states at different energy levels. (b)Stackstructure of DMTJ and setup of measurement.

FIG. 2 (Color on line). Normalized conductance as a function of bias voltage in (a) P and (b) AP state, respectively.(c)Differential TMR (DTMR) dependence with bias voltage.(d)QW thickness dependence of $d^2I/dV^2$ curves in P state (measured at 16K). The dash lines indicate the resonant peak positions. (e)QW peak positions for the experimental results (circle), PAM simulation results (lines) and *ab-initio* calculation results (square). The three numbers *n* in the figure representthe QW node number just below $E_F$ for the three samples with different QW thickness.

FIG.3(Color on line).$d^2I/dV^2$ curves in P and AP states for DMTJs with (a) MgO(3ML)/Fe/MgAlO$_x$(12ML) and (b) MgO(3ML)/Fe/MgO(12ML) structures, respectively. Insets:schematics of tunneling of electrons with $\Delta_1$ symmetry in AP configuration in samples with thick MgAlO$_x$ and MgO barrier, respectively. (c)Electrostatic potentialsand structuresof Fe/MgO and Fe/MgAl$_2$O$_4$ layers.The width (unit is Å) between potential valleys in Fe layers are shown for Fe/MgO (black) and Fe/MgAl$_2$O$_4$ (red) interface. The lattice constant and length of Fe-O bond are also shown in the structure.(d)Schematics of DOS in QW with large and small interface phase shift distribution. $\Phi_1$, $\Phi_2$ represent the phase shift on reflection at interface and $\Phi_{inf}$ stands for the interface phase shift.

**Table 1.** Summary of experimental results for samples with different barrier configurations.

| No. | Barrier I (3ML) | Barrier II (12ML) | TMR (10mV) | | Oscillation amplitude(a.u.) | | Fe QW $t$ (nm) |
|---|---|---|---|---|---|---|---|
| | | | T=295K | T=16K | P | AP | |
| A | MgAlO$_x$ | MgO | 192% | 297% | 0.86 | -- | 9.8 |
| B | MgO | MgO | 176% | 300% | 0.71 | -- | 9.6 |
| C | MgAlO$_x$ | MgAlO$_x$ | 130% | 177% | 6.55 | 1.72 | 12.6 |
| D | MgO | MgAlO$_x$ | 121% | 198% | 10.17 | 2.42 | 10.0 |

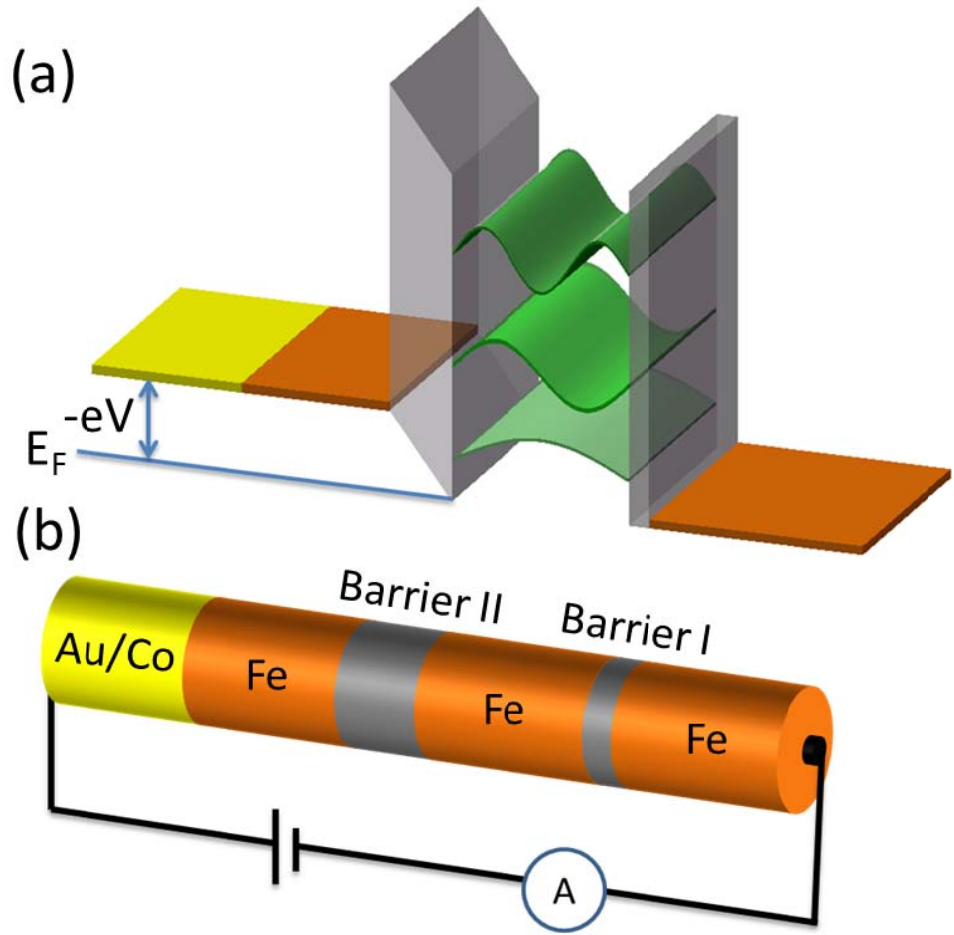

FIG1

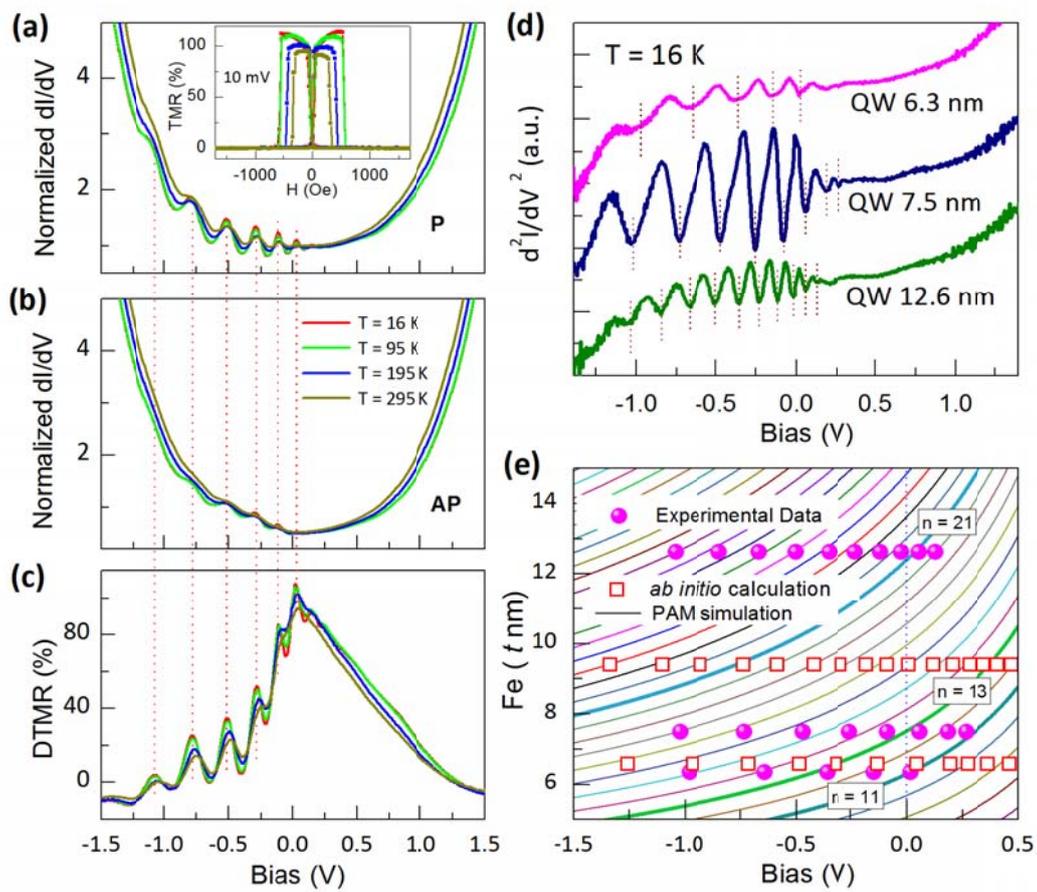

FIG2

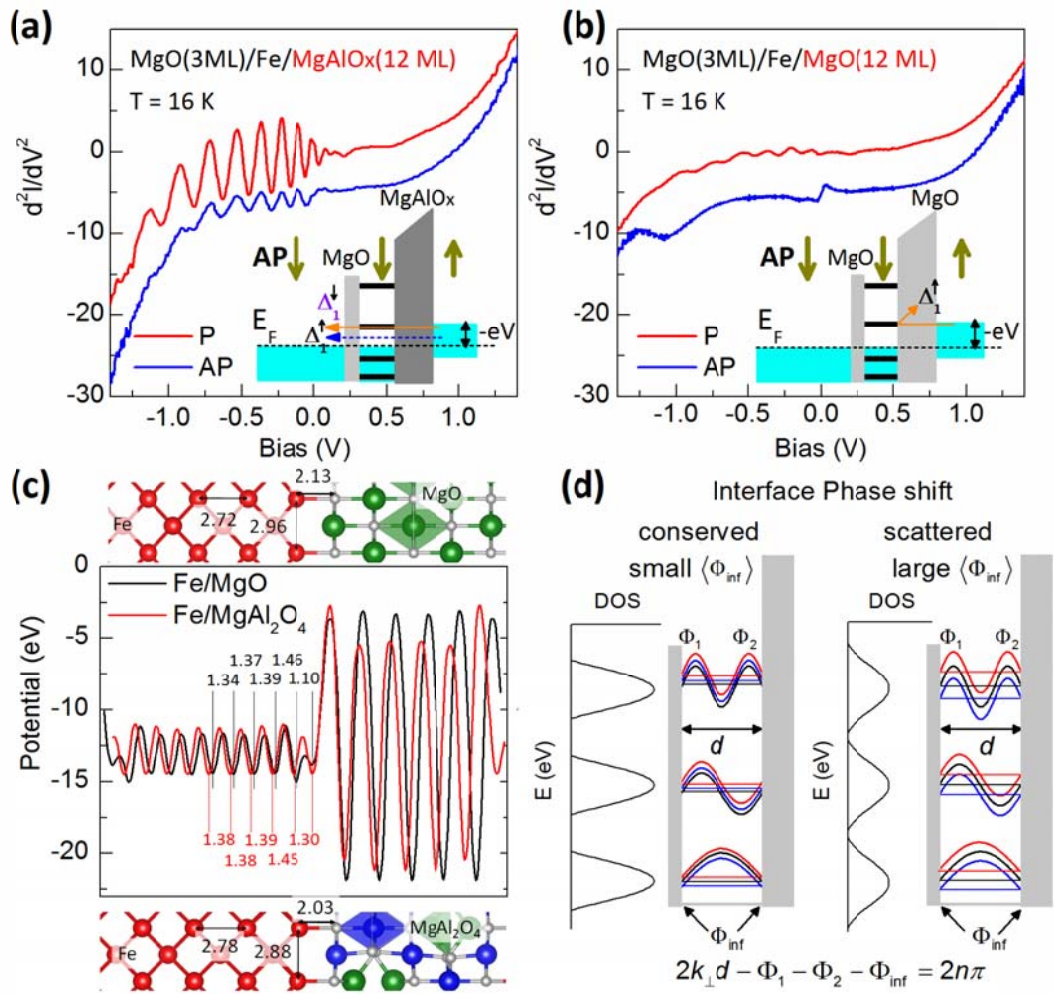

FIG3